\documentstyle [12pt]{article}

\def\b{\begin{equation}} \def\e{\end{equation}}
\def\bd{\begin{displaystyle}} \def\ed{\end{displaystyle}}
\def\ba{\begin{array}} \def\ea{\end{array}}

\def\bee{\begin{enumerate}}
\def\eee{\end{enumerate}}

\def\bes{\begin{eqnarray*}}
\def\ees{\end{eqnarray*}}
\def\be{\begin{eqnarray}}
\def\ee{\end{eqnarray}}

\begin{document}
\title{ QED effective action in Krein space quantization}

\author{A. Refaei$^{1,2}$\thanks{e-mail:
refaei@iausdj.ac.ir}, M.V. Takook$^1$\thanks{e-mail:
takook@razi.ac.ir}}

\maketitle \centerline{\it $^1$Department of Physics, Razi
University, Kermanshah, Iran} \centerline{\it $^2$Department of Physics,  Sanandaj branch, Islamic Azad
University, Sanandaj, Iran.}

\begin{abstract}
The one-loop effective action of QED is calculated by the
Schwinger method in Krein space quantization. We show that the
effective action  is naturally finite and regularized. It also
coincides with the renormalized solution which was derived by
Schwinger.

\end{abstract}
\emph{Keywords:} Krein space, Effective action

\section {Introduction}

The minimally coupled scalar field in de Sitter space plays an important role in the inflationary model as well as
in the linear quantum gravity. In 1985, Allen showed that a covariant quantization of minimally coupled scalar field cannot be constructed from positive
norm states alone \cite{allen85}. For obtaining a covariant quantization of this field, a new method of field quantization has been presented, {\it i.e.} Krein space quantization \cite{gazeau00}. It has been proven that the use of the two sets of solutions (positive and negative norm states) are an unavoidable feature for preservation
of (1) causality (locality), (2) covariance, and (3) elimination of the infrared divergence for the minimally coupled scalar field in de Sitter space. The most interesting result of this construction is the convergence of the Green function at large distances, which means that the infrared divergence is gauge dependent \cite{ta3}.

It has been shown that quantization in Krein space
removes all ultraviolet divergences of QFT except the light cone
singularity \cite{gazeau00,ta3,ta4}. It was conjectured that quantum metric fluctuations might smear out the singularities of Green functions on the light cone, but it does not remove other ultraviolet divergences \cite{ford97}. However, by using the Krein space quantization and
the quantum metric fluctuations in the linear approximation, we
showed that the problem of infinities in QFT disappears \cite{rota}.

This method was applied to different problem and the natural regularized results were
obtained \cite{gagarota,derotata}. We have computed the scalar field effective action in Krein space quantization which includes quantum metric fluctuation \cite{reta}.
A finite result is obtained naturally with the same physical results as yielded by the standard method. We hope this method can solve the problem of non-renormalizability of quantum gravity in the linear approximation.

In this paper, we address the problem of derivation of the
low-energy effective action, which is solved perturbatively for QED
\cite{j}. Using the Krein space method and quantum metric
fluctuation at the linear approximation, we calculate the one loop
effective action for QED. We have shown that the result not only
is regularized but also is equivalent to the renormalized result
which was reported by Schwinger.

The paper is organized as follows. In section 2, we briefly recall the propagator derivation for scalar field in Krein space
quantization. Section 3 is devoted to express the one-loop
effective action in terms of the new propagator in our method.
Then, we develop the Schwinger technique for
generating the perturbative expression in section 4. In this section we present
the main results of our paper. In appendices, we have provided some
calculation techniques which are needed.

\setcounter{equation}{0}
\section{Scalar Green function}

We review the elementary facts about Krein space
quantization. A classical scalar field $\phi(x)$ satisfies the
following field equation  \b \label{eq:1}
(\Box+m^2)\phi(x)=0=(\eta^{\mu\nu}\partial_\mu
\partial_\nu+m^2)\phi(x),\;\;
\eta^{\mu\nu}=\mbox{diag}(1,-1,-1,-1).\e
 Inner ({\it Klein-Gordon}) product and related norms are defined by \cite{bida}
\b(\phi_1,\phi_2)=-i\int_{t=\mbox{const.}}\phi_1(x)\stackrel{\leftrightarrow}
{\partial}_t\phi_2^*(x)d^3x.\e Two sets of solutions are given by:
\b u_p(k,x)=\frac{e^{i\vec k.\vec x-iwt}}{\sqrt{(2\pi)^32w}}
=\frac{e^{-ik.x}}{\sqrt{(2\pi)^32w}},\e
\b u_n(k,x)=\frac{e^{-i\vec
k.\vec x+iwt}}{\sqrt{(2\pi)^32w}}
=\frac{e^{ik.x}}{\sqrt{(2\pi)^32w}},\e where $ w(\vec k)=k^0=(\vec
k .\vec k+m^2)^{\frac{1}{2}} \geq 0$, note that $u_n$ has the
negative norm. In Krein space the quantum field is defined as
follows \cite{ta4}:
 \b \phi(x)=\frac{1}{\sqrt
2}[\phi_p(x)+\phi_n(x)],\e
where $$ \phi_p(x)=\int d^3\vec k [a(\vec k)u_p(k,x)+a^{\dag}(\vec
k)u_p^*(k,x)],$$ $$ \phi_n(x)=\int d^3\vec k [b(\vec
k)u_n(k,x)+b^{\dag}(\vec k)u_n^*(k,x)].$$  $a(\vec k)$ and $b(\vec
k)$ are two independent operators. The time-ordered product
propagator for this field operator is \b iG_T(x,x')=<0\mid
T\phi(x)\phi(x') \mid 0>=\theta (t-t'){\cal W}(x,x')+\theta
(t'-t){\cal W}(x',x).\e In this case we obtain \b
G_T(x,x')=\frac{1}{2}[G_F(x,x')+(G_F(x,x'))^*]=\Re G_F(x,x'),\e
where the Feynman Green function is defined by \cite{bida}
\begin{eqnarray}
G_F(x,x')&=&\int \frac{d^4 p}{(2\pi)^4}e^{-ip.(x-x') }\tilde
G_F(p)=\int \frac{d^4p}{(2\pi)^4}\frac{e^{-ip.(x-x')}}{p^2-m^2+i\epsilon} \label{eq:2}
\nonumber\\&=&-\frac{1}{8\pi}\delta
(\sigma_0)+\frac{m^2}{8\pi}\theta(\sigma_0)\frac{J_1
(\sqrt{2m^2\sigma_0})-iN_1 (\sqrt{2m^2\sigma_0})}{\sqrt{2m^2
\sigma_0}}\nonumber\\ & & -\frac{im^2}{4\pi^2}\theta(-\sigma_0)\frac{K_1
(\sqrt{-2m^2\sigma_0})}{\sqrt{-2m^2 \sigma_0}},\end{eqnarray}
 where
$\sigma_0=\frac{1}{2}(x-x')^2 .$ So we have \begin{eqnarray} G_T(x,x')&=&\int
\frac{d^4 p}{(2\pi)^4}e^{-ip.(x-x')}{\cal
PP}\frac{1}{p^2-m^2}\nonumber\\ &=&-\frac{1}{8\pi}\delta
(\sigma_0)+\frac{m^2}{8\pi}\theta(\sigma_0)\frac{J_1
(\sqrt{2m^2\sigma_0})}{\sqrt{2m^2 \sigma_0}}, \;\;x\neq x',
\label{28}
\end{eqnarray}
${\cal PP}$ stands for the principal parts. Contribution of the
coincident point singularity $(x=x')$ merely appears in the
imaginary part of $G_F$ (\cite{ta3} and equation (9.52) in \cite{bida})
$$ G_F(x,x)=-\frac{2i}{(4\pi)^2}\frac{m^2}{d-4}+G_F^{\mbox{finit}}(x,x),$$ where $d$ is the space-time dimension and
$G_F^{\mbox{finit}}(x,x)$ becomes finite as $d\longrightarrow 4$.
Note that the singularity of the Eq.(\ref{28}) takes place only on
the cone \emph{i.e.,} $x\neq x', \sigma_0=0$.

It has been shown that the quantum metric fluctuations remove the
singularities of Green's functions on the light cone \cite{ford97}.
Therefore, the quantum field theory in Krein space, including the
quantum metric fluctuation $\left(
g_{\mu\nu}=\eta_{\mu\nu}+h_{\mu\nu}\right)$, removes all the
ultraviolet divergencies of the theory \cite{rota,ford97}, so one
can write: \b \langle G_T(x,x')\rangle = -\frac{1 }{8\pi}
\sqrt{\frac{\pi}{2\langle\sigma_1^2\rangle}}
exp\left(-\frac{\sigma_0^2}{2\langle\sigma_1^2\rangle}\right)+
 \frac{m^2}{8\pi}\theta(\sigma_0)\frac{J_1(\sqrt {2m^2
 \sigma_0})}{\sqrt {2m^2 \sigma_0}},\label{210}\e
where $2\sigma= g_{\mu\nu}(x^{\mu}-
x'^{\mu})(x^{\nu} - x'^{\nu})$ and $\sigma_1$ is the first order shift in $\sigma$, due to
the linear quantum gravity ($\sigma=\sigma_0+\sigma_1+\emph{O}(h^2)$). The average value is taken over the quantum metric fluctuation and in the case of $2\sigma_0 =\eta_{\mu\nu}(x^{\mu}- x'^{\mu})(x^{\nu} - x'^{\nu})=0$ we have
$\langle\sigma_1^2\rangle\neq 0$. So, we get \b \langle
G_T(0)\rangle = -\frac{1 }{8\pi}
\sqrt{\frac{\pi}{2\langle\sigma_1^2\rangle}} +
 \frac{m^2}{8\pi}\frac{1}{2}.\e
It should be noted that $ \langle\sigma_1^2\rangle $ is related to
the density of gravitons \cite{ford97}.

By using the Fourier transformation of Dirac delta function,
$$ -\frac{1}{8\pi}\delta(\sigma_0)= \int \frac{d^4
p}{(2\pi)^4}e^{-ip.(x-x')}{\cal PP}\frac{1}{p^2},$$or equivalently
$$\frac{1}{8\pi^2}\frac{1}{\sigma_0}= -\int \frac{d^4
p}{(2\pi)^4}e^{-ip.(x-x')}\pi\delta(p^2),$$ for the second part of
Green function, we obtain \b
\frac{m^2}{8\pi}\theta(\sigma_0)\frac{J_1(\sqrt {2m^2
 \sigma_0})}{\sqrt {2m^2 \sigma_0}}=\int \frac{d^4
p}{(2\pi)^4}e^{-ip.(x-x')}{\cal PP}\frac{m^2}{p^2(p^2-m^2)}.\e
And for the first part we have $$
-\frac{1}{8\pi}\sqrt{\frac{\pi}{2\langle\sigma_1^2\rangle}}exp\left[-\frac{(x-x')^4}
{4\langle\sigma_1^2\rangle}\right]=\int \frac{d^4
p}{(2\pi)^4}e^{-ik.(x-x') }  \tilde{G}_1(p),$$
 where $\tilde{G_1}$ is fourier transformation of the first part of the Green function (\ref{210}). Therefore, we
obtain \b \label{eq:22}<\tilde G_T(p)>=\tilde{G}_1(p)+{\cal
PP}\frac{m^2}{p^2(p^2-m^2)}.\e In the previous paper, we proved
that in the one-loop approximation, the Green function in Krein
space quantization, which appears in the transition amplitude is
\cite{ta4}: \b <\tilde G_T(p)>\mid_{\mbox{one-loop}}\equiv \tilde
G_T(p)\mid_{\mbox{one-loop}}\equiv {\cal PP}
\frac{m^2}{p^2(p^2-m^2)}.
\label{213} \e
 That means in the one loop
approximation, the contribution of $\tilde{G_1}(p)$ is negligible.
It is worth to mention that in order to improve the UV behavior in
relativistic higher-derivative correction theories, the propagator
(\ref{213}) has been used by some authors \cite{bach,ho}. It is also appear in supersymmetry (equation (20.76) in \cite{ka}).

\setcounter{equation}{0}
\section{One-loop effective action}

Let us start from the general method which was originally
developed in \cite{reta}. The one-loop effective action in QED
reduces to computing the fermion determinant
\begin{eqnarray} J&=& \frac{i}{2} Tr
\ln \left[1-\frac{2eA.p+\frac{e}{2}\sigma_{\mu\nu}
F^{\mu\nu}-e^2A^2}{p^2-m^2+i\epsilon}\right] \nonumber\\
&=&\frac{i}{2}
\int d^4x<x|\ln \left[1-\frac{2eA.p+\frac{e}{2}\sigma_{\mu\nu}
F^{\mu\nu}-e^2A^2}{p^2-m^2+i\epsilon}\right]|x> \label{315}.\end{eqnarray}
 One can write this determinant, in proper time method, as
\cite{itzu}
\begin{eqnarray}
J&=&\frac{i}{2}\int_{0}^{\infty}dss^{-1}e^{-ism^2}Tr\exp\left\{-is[p^2-e(p.A+A.p)-\frac{e}{2}\sigma_{\mu\nu}F^{\mu\nu}+e^2A^2]\right\}\nonumber\\
&-& \frac{i}{2}\int_{0}^{\infty}dss^{-1}e^{-ism^2}Tr \exp(-isp^2)\nonumber\\
&=&\frac{i}{2}\int_{0}^{\infty}dss^{-1}e^{-ism^2}\left[Tr U(s)- Tr
U_0(s)\right], \end{eqnarray}
 where
$U(s)=\exp\left\{-is[p^2-e(p.A+A.p)-\frac{e}{2}\sigma_{\mu\nu}F^{\mu\nu}+e^2A^2]\right\}$
and $U_0(s)=\exp(-isp^2).$ In Krein space quantization including
the quantum metric fluctuation, equation $(3.1)$ reads as \b
J_{kr}=\frac{i}{2} Tr \ln \left[1-(2eA.p+
\frac{e}{2}\sigma_{\mu\nu} F^{\mu\nu}-e^2A^2) {\cal PP}
\frac{m^2}{p^2(p^2-m^2)}\right].\e If we
take$$V=\frac{1}{2}m^2(2eA.p+\frac{e}{2}\sigma_{\mu\nu}
F^{\mu\nu}-e^2A^2),$$ we can write
\begin{eqnarray}J_{kr}&=&\frac{i}{2}Tr\ln
\Big[1- {\cal PP} \frac{2V}{p^2(p^2-m^2)}\Big]\nonumber\\
&=&\frac{i}{2}Tr\ln \Big[1-
V\Big(\frac{1}{p^2(p^2-m^2)+i\epsilon}+\frac{1}{p^2(p^2-m^2)-i\epsilon}\Big)\Big]\nonumber\\
&=&\frac{i}{2} Tr\ln \Big[\Big(1-\frac{V}{p^2(p^2-m^2)-i\epsilon
}\Big)\Big(1-\frac{V}{p^2(p^2-m^2)+i\epsilon
}\Big)-\Big(\frac{V}{p^2(p^2-m^2)}\Big)^2 \Big],
\label{317} \end{eqnarray}
where $\epsilon^2$ has been vanished. By continuing
the calculation, for this equation we obtain
\begin{eqnarray} J_{kr}&=&\frac{i}{2}Tr\ln \Big[
\Big(1-\frac{V}{p^2(p^2-m^2)-i\epsilon }\Big)
\Big(1-\frac{V}{p^2(p^2-m^2)+i\epsilon }\Big)\nonumber\\
& &\times \Big(1-\frac{(\frac{V}{p^2(p^2-m^2)})^2}{(1-\frac{V}{p^2(p^2-m^2)
 -i\epsilon})(1-\frac{V}{p^2(p^2-m^2)+i\epsilon })}\Big)\Big]\nonumber\\
&=&\frac{i}{2}Tr\ln\Big(1-\frac{V}{p^2(p^2-m^2)-i\epsilon
}\Big)+\frac{i}{2}Tr \ln\Big(1-\frac{V}{p^2(p^2-m^2)+i\epsilon
}\Big)\nonumber\\
& &+\frac{i}{2}Tr\ln\Big[1-\Big(\frac{(\frac{V}{p^2(p^2-m^2)})^2}
{(1-\frac{V}{p^2(p^2-m^2)-i\epsilon
})(1-\frac{V}{p^2(p^2-m^2)+i\epsilon })}\Big)\Big] .\end{eqnarray}  The
last term in this equation splits into two terms. So, $J_{kr}$
becomes \begin{eqnarray} J_{kr}&=&
\frac{i}{2}Tr\ln\left(1-\frac{V}{p^2(p^2-m^2)-i\epsilon
}\right)+\frac{i}{2}Tr \ln \left(1-\frac{V}{p^2(p^2-m^2)+i\epsilon
}\right)\nonumber\\ & &+\frac{i}{2}
Tr\ln\left(1-\frac{V}{p^2(p^2-m^2)-V}\right) + \frac{i}{2}
Tr\ln\left(1+\frac{V}{p^2(p^2-m^2)-V}\right).\end{eqnarray} By using the
results in appendix B, we obtain:
  \begin{eqnarray} J_{kr}&=&\frac{i}{2}Tr\ln\left(1+\frac{Y}{p^2}\right)+\frac{i}{2}Tr\ln\left(1-\frac{Y}{p^2-m^2}\right)\nonumber\\
& &+\frac{i}{2}Tr\ln\left(1+\frac{Y^2}{(m^2p^2+Y)(m^2p^2-m^4-Y)}\right)
, \end{eqnarray}
where $V=\frac{m^2 Y}{2}$ and
$ Y =2eA.P+\frac{e}{2}\sigma_{\mu\nu} F^{\mu\nu}-e^2A^2$. We
define the functions $J$, $J_0$ and $J_1$ as follow:
$$J=\frac{i}{2}Tr\ln\left(1-\frac{Y}{p^2-m^2}\right) ,\qquad J_{0}=\frac{i}{2}Tr\ln\left(1+\frac{Y}{p^2}\right)\\, $$
$$ J_{1}=\frac{i}{2}Tr\ln\left(1+\frac{Y^2}{(m^2p^2+ Y )(m^2p^2-m^4- Y )}\right),$$
so, we  have
\b J_{kr}=J+J_{0}+J_{1}.\label{eq39}\e

\setcounter{equation}{0}
\section{Regularized effective action}

The approximate evaluation of $J_{kr}$ is the main purpose of this
article. Therefore, following the perturbation approach developed
in \cite{j}, we discuss the evaluations of $J$, $J_{0}$ and $
J_{1}$ in equation (\ref{eq39}), which lead to derive the former
expression for the one loop effective action in an external field.
It's no difficult to show that
\begin{eqnarray}J&=&\frac{i}{2}Tr\ln\left(1-\frac{2eA.P+\frac{e}{2}\sigma_{\mu\nu}
F^{\mu\nu}-e^2A^2}{p^2-m^2+i\epsilon}\right)\nonumber\\&=&\frac{i}{2}\int_{0}^{\infty}dss^{-1}e^{-ism^2}\left[Tr
U(s) - Tr U_0(s)\right]\nonumber\\
&=&W^{(1)}-\frac{i}{2}\int_{0}^{\infty}dss^{-1}e^{-ism^2}Tr
U_0(s)\nonumber\\&=&-\frac{e^2}{12\pi^2}\int_0^{\infty}dss^{-1}\exp(-m^2s)\int
d^4k\frac{1}{4}F_{\mu\nu}(-k)F_{\mu\nu}(k)\nonumber\\&&+
 \frac{e^2}{4\pi^2}\int d^4k\frac{k^2}{4}F_{\mu\nu}(-k)F_{\mu\nu}(k)\int_0^1 dv \frac{v^2(1-\frac{1}{3}v^2)}{m^2+\frac{k^2}{4}(1-v^2)},\end{eqnarray}
and
\begin{eqnarray}J_0&=&\frac{i}{2}Tr\ln\left(1+\frac{2eA.P+\frac{e}{2}\sigma_{\mu\nu}
F^{\mu\nu}-e^2A^2}{p^2+i\epsilon}\right)\nonumber\\&=&-\frac{e^2}{4\pi^2}\int_0^{\infty} dss^{-2} \int d^4k A_{\mu}(-k)A_{\mu}(k)\nonumber\\&&-\frac{e^2}{12\pi^2}\int_0^{\infty}dss^{-1}\int d^4k\frac{1}{4}F_{\mu\nu}(-k)F_{\mu\nu}(k)\nonumber\\&&+
 \frac{e^2}{16\pi^2}\int d^4kF_{\mu\nu}(-k)F_{\mu\nu}(k)\int_0^1 v^2dv \frac{1-\frac{1}{3}v^2}{1-v^2}.\end{eqnarray}
We have employed the methods which are presented in appendix A. In $J_1$,
we expand the logarithm function and keep the first term
\begin{eqnarray}J_1&=&\frac{i}{2}Tr\ln\left(1+\frac{Y^2}{(p^2+Y)(p^2-m^2-Y)}\right)\simeq \frac{i}{2}Tr\frac{Y^2}{(p^2+Y)(p^2-m^2-Y)}\nonumber\\
&=&-\frac{i}{2}Tr\frac{Y^2}{m^2+2Y}\left(\frac{1}{p^2+Y}-
\frac{1}{p^2-m^2-Y}\right),\end{eqnarray}
 and so we can write
\begin{eqnarray}J_1&=&-\frac{i}{2}Tr\frac{Y^2}{i(m^2+2Y)}\int_0^\infty dse^{-isp^2}\left(e^{-isY}-e^{is(Y+m^2)}\right),\nonumber\\
&=&-\frac{i}{2}Tr\frac{Y^2}{im^2}\left[1+\sum_{n=1}\left(-\frac{2Y}{m^2}\right)^n\right]\int_0^\infty
ds e^{-isp^2}\left( e^{-isY}-e^{is(Y+m^2)}\right).\end{eqnarray}
 We
restrict ourselves to a specific finite number of the powers of
$Y$ to compare with the expressions of $J$ and $J_{0}$, hence, we
have \b J_1\simeq \frac{i}{2}Tr\int_0^\infty s ds
Y^2e^{-isp^2}.\e One can write [Appendix A]:
\begin{eqnarray}J_1&\simeq & \frac{ie^2}{2}\int_0^{\infty} s ds \nonumber\\ & & \times\Big(\frac{1}{2} \int_{-1}^{1}dv Tr\Big[(p.A+A.p) \exp\Big(-ip^2\frac{1}{2}(1-v)s\Big) \times
 (p.A+A.p)\exp\Big(-ip^2\frac{1}{2}(1+v)s\Big)\Big]\nonumber\\ & &
 + \frac{1}{2}\int_{-1}^{1}dv Tr\Big[\frac{1}{2}\sigma F
 \exp\Big(-ip^2\frac{1}{2}(1-v)s\Big)\times \frac{1}{2}\sigma F  \exp\Big(-ip^2\frac{1}{2}(1+v)s\Big)\Big] \Big).\end{eqnarray}
 Now, we calculate these traces in a momentum representation
 \begin{eqnarray}J_1&\simeq & \frac{2ie^2}{(2\pi)^4}\int_0^{\infty} sds \times  \Big(\frac{1}{2}\int_{-1}^{1}dv \int d^4k\int d^4p 2p.A(-k)
 \exp\Big[-i\Big(p+\frac{1}{2}k\Big)^2\frac{1}{2}(1-v)s\Big]\nonumber\\ & &\times
 2p.A(k) \exp\Big[-i\Big(p-\frac{1}{2}k\Big)^2\frac{1}{2}(1+v)s\Big]
 +\frac{1}{2}\int_{-1}^{1}dv \int d^4k \int d^4p \frac{1}{4} tr\frac{1}{2}\sigma F\nonumber\\ & &
 \times\exp\Big[-i\Big(p+\frac{1}{2}k\Big)^2\frac{1}{2}(1-v)s\Big] \frac{1}{2}\sigma F
 \exp\Big[-i\Big(p-\frac{1}{2}k\Big)^2\frac{1}{2}(1+v)s\Big]\Big),\end{eqnarray}
 then
\begin{eqnarray}J_1&\simeq & \frac{2ie^2}{(2\pi)^4}\int_0^{\infty} s ds \Big[(-2\pi^2s^{-3})\int d^4k
A_{\mu}(-k)A_{\mu}(k)\nonumber\\ & &- \int(i\pi^2s^{-2}) d^4k\frac{1}{2}F_{\mu\nu}(-k)F_{\mu\nu}(k)\int_0^1 dv(1-v^2)e^{-\frac{ik^2(1-v^2)s}{4}} \Big].\end{eqnarray}
 Finally, we obtain $J_1$ in the first order approximation as
   \begin{eqnarray}J_1&\simeq & \frac{e^2}{4\pi^2}\int_0^{\infty} s^{-2} ds \int d^4k A_{\mu}(-k)A_{\mu}(k)\nonumber\\ & &+\frac{e^2}{6\pi^2}\int_0^{\infty} s^{-1}ds \int d^4k\frac{1}{4}F_{\mu\nu}(-k)F_{\mu\nu}(k)\nonumber\\ & &-\frac{e^2}{8\pi^2}\int d^4kF_{\mu\nu}(-k)F_{\mu\nu}(k) \int_0^1 dv
\frac{v^2(1-\frac{1}{3}v^2)}{1-v^2}.\end{eqnarray}
 By replacing $J$, $J_{0}$ and $ J_{1}$ in equation Eq.(\ref{eq39}) we find
 $$J_{kr}=J+J_0+J_1$$
 \begin{eqnarray} J_{kr}&=&-\frac{e^2}{12\pi^2}\int_0^{\infty}dss^{-1}\exp(-m^2s) \int d^4k\frac{1}{4}F_{\mu\nu}(-k)F_{\mu\nu}(k)\nonumber\\ & &+
 \frac{e^2}{4\pi^2}\int d^4k\frac{k^2}{4}F_{\mu\nu}(-k)F_{\mu\nu}(k)\int_0^1 dv \frac{v^2(1-\frac{1}{3}v^2)}{m^2+\frac{k^2}{4}(1-v^2)}\nonumber\\ & &
 +\frac{e^2}{12\pi^2}\int_0^{\infty} s^{-1}ds \int d^4k\frac{1}{4}F_{\mu\nu}(-k)F_{\mu\nu}(k)\nonumber\\ & &-\frac{e^2}{16\pi^2}\int d^4kF_{\mu\nu}(-k)F_{\mu\nu}(k) \int_0^1 dv \frac{v^2(1-\frac{1}{3}v^2)}{1-v^2}.
\end{eqnarray}
This equation can be written in the following form: \b
J_{kr}=\frac{e^2}{16\pi^2}\int d^4kF_{\mu\nu}(-k)F_{\mu\nu}(k)
  \left[I_1 + I_2 +I_3+I_4\right], \e
where
$$ I_1=-\frac{1}{3}\int_0^{\infty}dss^{-1}\exp(-m^2s),\;\;\;\;I_3=\frac{1}{3}\int_0^{\infty}
s^{-1}ds,$$
$$I_2=\int_0^1 dv
k^2\frac{v^2(1-\frac{1}{3}v^2)}{m^2+\frac{k^2}{4}(1-v^2)},\;\;\;
I_4=-\int_0^1 dv \frac{v^2(1-\frac{1}{3}v^2)}{1-v^2}.$$ The
integrals $I_1$, $I_3$ and $I_4$ are divergence. $I_1$ is:
$$I_1=-\frac{1}{3}\int_0^{\infty}dss^{-1}\exp(-m^2s)=-\frac{1}{3}\Gamma(0).$$
By using the following relations:
 $$\lim_{x \rightarrow
0}\Gamma(x)=\lim_{x \rightarrow 0} E_{1}(x),$$
$$E_{1}(x)=\int_{x}^{\infty}\frac{e^{-t}}{t}dt=-\gamma - \ln x - \sum_{n=1}\frac{(-1)^n x^n}{nn!},$$ where $\gamma$ is the Euler's
constant, we can write
 $$ \Gamma(0)=-\gamma-\lim_{x\rightarrow 0} \ln x ,$$
or
$$I_1=\frac{1}{3}(\gamma+\lim_{x\rightarrow 0} \ln x ).$$
The divergence form of $I_3$ is
$$ I_3=\frac{1}{3}\int_0^{\infty} s^{-1}ds=\frac{1}{3}\left(\lim_{\Lambda\rightarrow\infty} \ln{\Lambda}-\lim_{\mu\rightarrow 0}\ln{\mu}\right).$$
 The $I_4$ divergency, which has been discussed in Appendix B, is as:
\b I_4=-\int_0^1 dv \frac{v^2(1-\frac{1}{3}v^2)}{1-v^2}= \frac{1}{3}\lim_{\mu\rightarrow0}\ln{\mu}-\frac{1}{3}\ln2+\frac{5}{9}. \e
Therefore, $J_{kr}$ becomes: \begin{eqnarray}J_{kr}&=&\frac{e^2}{16\pi^2}\int
d^4k F_{\mu\nu}(-k)F_{\mu\nu}(k)\times
  \Big[\frac{1}{3}(\gamma+\lim_{x\rightarrow 0} \ln x )+\frac{1}{3}\Big(\lim_{\Lambda\rightarrow\infty} \ln{\Lambda}
   -\lim_{\mu\rightarrow 0}\ln{\mu}\Big)\nonumber\\ & &-\frac{1}{3}\Big( -\lim_{\mu\rightarrow0}\ln{\mu}+\ln2
  -\frac{5}{3} \Big)+\int_0^1 dv k^2\frac{v^2(1-\frac{1}{3}v^2)}{m^2+\frac{k^2}{4}(1-v^2)}\Big].
\end{eqnarray}
 It's well known that $$\lim_{\Lambda\rightarrow\infty} \ln{\Lambda}\equiv-\lim _{x\rightarrow 0}
 \ln x$$ then, we see that $W_{kr}=W^{0}+J_{kr}$ reduces to
\begin{eqnarray} W_{kr}&=&-\int d^4k\frac{1}{4}F_{\mu\nu}(-k)F_{\mu\nu}(k)\nonumber\\ & &\times\left[1-
\frac{\alpha}{4\pi}\left(\frac{k^2}{m^2}\int_0^1 dv \frac{v^2(1-\frac{1}{3}v^2)}{1+\frac{k^2}{4m^2}(1-v^2)}-\frac{4}{3}\left(\ln2-\frac{5}{3}-\gamma\right)\right)\right],\end{eqnarray}
 where we have added the action integral of the Maxwell field, which is expressed in momentum space by
  $$W^{(0)}=-\int d^4k\frac{1}{4}F_{\mu\nu}(-k)F_{\mu\nu}(k),$$
and we took $\alpha=\frac{e^2}{4\pi}$.
If we put $\alpha=\frac{1}{137}$ and $\gamma=0.5772156649...$,
finally, the total effective action is
\begin{eqnarray}W_{kr}&=&-\int d^4k\frac{1}{4}F_{\mu\nu}(-k)F_{\mu\nu}(k)\nonumber\\ & &\times\left[0.9987989919- \frac{\alpha}{4\pi}\frac{k^2}{m^2}\int_0^1 dv \frac{v^2(1-\frac{1}{3}v^2)}{1+\frac{k^2}{4m^2}(1-v^2)}\right].\end{eqnarray}
 Now, one compares this result with standard solution Eq.(\ref{eq:b}).

This new kind of regularization may be utilized in the calculation
of the Lamb-Shift and Magnetic-Anomaly \cite{zafota}.

\section{Conclusion}
We recall that the negative frequency solutions of the field equation is needed for quantizing the minimally coupled scalar field in de Sitter space. Contrary to the
Minkowski space, the elimination of de Sitter negative norms in the minimally coupled states
breaks the de Sitter invariance. Then, for restoring the de Sitter invariance, one needs to take
into account the negative norm states i.e. the Krein space quantization. It provides a natural
tool for eliminating the singularity in the QFT.

Here, it is found that in this approximation the theory is free of
any divergence and the effective action coincides with
standard solution. So, for QED, we see that this quantization
eliminates the singularity in the theory without changing the
physical content of the theory in the one-loop approximation. This method can be used as an alternative way for solving the non-renormalizability of quantum gravity in the linear approximation.

\vskip 0.5 cm

\noindent {\bf{Acknowledgments}}: The authors would like to thank S. Rouhani.

\begin{appendix}
\setcounter{equation}{0}
 \section{Schwinger method}

In this appendix, in order to make the paper self-contained, we will present all of Schwinger's method as a
major reference for our calculation.
We discuss the approximate
evaluation of
$$W^{(1)}=i\frac{1}{2}\int_{0}^{\infty}dss^{-1}exp{(-im^2s)}TrU(s),$$
by an expansion in powers of $eA_{\mu}$ and $eF_{\mu\nu}$. So, we can write
$$ H=H_{0}+H_{1},$$
where
$$H_0=p^2 ,\qquad H_1=-e(p.A+A.p)-\frac{e}{2}\sigma_{\mu\nu}F^{\mu\nu}+e^2A^2.$$
To obtain the expansion of $Tr U(s)$ in powers of $H_1$, we
observe that $U(s)$ obeys the differential equation
$$i\partial_sU(s)=(H_0+H_1)U(s).$$
The related operator

$$V(s)=U_0^{-1}(s)U(s),$$
where$$U_{0}=exp(-iH_0s),$$
is determined by
\b i\partial_s V(s)=U_0^{-1}(s)H_1U_0(s)V(s),\e
and
$$V(0)=1.$$
From $Eq.(A.1)$ one can obtain
$$V(s)=1-i\int_{0}^{s}ds'U_0^{-1}(s')H_1U_0(s')V(s'),$$
and construct the solution by iteration:
\b V(s)=1-i\int_{0}^{s}ds'U_0^{-1}(s')H_1U_0(s')+(-i)^2\int_{0}^{s}ds'U_0^{-1}(s')H_1U_0(s')\times\int_{0}^{s'}ds''U_0^{-1}(s'')H_1U_0(s'')+... .\e
On introducing new variables of integration, $u_1,u_2,...,$  according to $$ s'=su_1, s''=s'u_2,...,$$
 we obtain the expansion
 \b U(s)=exp(-iHs)=U_0(s)-is\int_{0}^{1}du_1U_0((1-u_1)s)H_1U_0(u_1s)+...
 $$$$(-is)^n\int_{0}^{1}u_1^{n-1}du_1...\int_0^1 du_nU_0((1-u_1)s)H_1U_0(u_1(1-u_1)s)...
 $$$$\times U_0(u_1u_2...u_{n-1}(1-u_n)s)H_1U_0(u_1u_2...u_ns)+... .\e
 Instead of taking the trace of this expression directly, which would involve further simplifications, we remark that

 $$TrU(s)-TrU_0(s)=-is\int_0^1d \lambda Tr[H_1 exp(-i(H_0+\lambda H_1)s)],$$
by the expansion of $ exp(-i(H_0+\lambda H_1)s)]$, one can write
 $$TrU(s)=TrU_0(s)+ (-is)Tr[H_1U_0(s)]+\frac{1}{2}(-is)^2\int_{0}^{1}du_1Tr[H_1U_0((1-u_1)s)H_1U_0(u_1s)+...
 $$$$\frac{(-is)^{n+1}}{n+1}\int_{0}^{1}u_1^{n-1}du_1...\int_0^1 du_nTr[H_1U_0((1-u_1)s)H_1U_0(u_1(1-u_1)s)...
 $$$$\times U_0(u_1u_2...u_{n-1}(1-u_n)s)H_1U_0(u_1u_2...u_ns)]+... $$
 We shall retain only the first nonvanishing field dependent terms in this expansion:
 \b W^{(1)}=\frac{1}{2}ie^2\int_0^{\infty} dss^{-1}\exp(-im^2s)\times \left\{ -is Tr[A^2exp(-ip^2s)]+ \right.$$$$ \left.
 \frac{1}{2}(-is)^2\int_{-1}^{1}\frac{1}{2}dv Tr\left[(pA+Ap) \exp\left(-ip^2\frac{1}{2}(1-v)s\right)\times (pA+Ap)\exp\left(-ip^2\frac{1}{2}(1+v)s\right)\right] \right.$$ $$ \left.
 +\frac{1}{2}(-is)^2\int_{-1}^{1}\frac{1}{2}dv Tr\left[\frac{1}{2}\sigma F  \exp \left(-ip^2\frac{1}{2}(1-v)s\right)\times \frac{1}{2}\sigma F  \exp(-ip^2\frac{1}{2}(1+v)s)\right] \right\}. \e

 For convenience, the variable $u_1$has been replaced by $\frac{1}{2}(1+v).$ The evaluation of these traces is naturally performed in a momentum representation. The matrix elements of the coordinate dependent field quantities depend only on momentum differences,
$$ \langle P+\frac{1}{2}k|A_{\mu}| P-\frac{1}{2}k\rangle=\frac{1}{(2\pi)^4}\int dx e^{-ikx}A_{\mu}(x)\equiv(2\pi)^{-2}A_{\mu}(k)$$ and
$$ \langle P|A^2_{\mu}| P\rangle=\frac{1}{(2\pi)^4}\int dxA^2_{\mu}(x)=(2\pi)^{-4}\int dk A_{\mu}(-k)A_{\mu}(k).$$
 Therefore
 \b W^{(1)}=\frac{2ie^2}{(2\pi)^4}\int_0^{\infty} dss^{-1}\exp(-im^2s)\times \left\{ -is\int d^4k A_{\mu}(-k)A_{\mu}(k)\int d^4p \exp(-ip^2s)+\right. $$$$\left.
 \frac{1}{2}(-is)^2\int_{-1}^{1}\frac{1}{2}dv \int d^4k\int d^4p 2p_{\mu}A_{\mu}(-k)\right.$$$$\left. \times \exp\left[-i\left(p+\frac{1}{2}k\right)^2\frac{1}{2}(1-v)s\right] 2p_{\nu}A_{\nu}(k)\exp\left[-i\left(p-\frac{1}{2}k\right)^2\frac{1}{2}(1+v)s\right]\right.$$$$\left.
 +\frac{1}{2}(-is)^2\int_{-1}^{1}\frac{1}{2}dv \int d^4k \int d^4p \frac{1}{4} tr\frac{1}{2}\sigma F  \right.$$$$\left. \times \exp\left(-i\left(p+\frac{1}{2}k\right)^2\frac{1}{2}(1-v)s\right) \frac{1}{2}\sigma F  \exp\left[-i\left(p-\frac{1}{2}k\right)^2\frac{1}{2}(1+v)s\right] \right\}.\e
 We thus encounter the elementary integrals
 $$ \int d^4p \exp(-ip^2s)=-i\pi^{2}s^{-2},$$
 $$ \int d^4p \exp\left[-i\left(p^2+\frac{k^2}{4}\right)s+ipkvs \right]=-i\pi^{2}s^{-2}\exp\left[-i\frac{k^2}{4}(1-v^2)s\right],$$
  $$ \int d^4p p_{\mu}p_{\nu} \exp\left[-i\left(p^2+\frac{k^2}{4}\right)s+ipkvs \right]=-i\pi^{2}s^{-2}\left(-\frac{i}{2}s^{-1}\delta_{\mu\nu}+ \frac{1}{4}v^2k_{\mu}k_{\nu}\right)\exp\left[-i\frac{k^2}{4}(1-v^2)s\right].$$

 It is convenient to replace the $\delta_{\mu\nu}$ term of the last integral by an expression which is equivalent to it in virtue of the integration with respect to $v$. Now
 $$\int_{-1}^{1}\frac{1}{2}dv\exp\left[-i\frac{k^2}{4}(1-v^2)s\right]=1-is\frac{1}{2}k^2\int_{-1}^{1}\frac{1}{3}dv v^2 \exp\left[-i\frac{k^2}{4}(1-v^2)s\right],$$
 so that, effectively
 $$ \int d^4p p_{\mu}p_{\nu} \exp\left[-i\left(p^2+\frac{k^2}{4}\right)s+ipkvs\right]=-\frac{1}{2}\pi^{2}s^{-3}\delta_{\mu\nu}$$$$+
 \frac{i}{4}\pi^2s^{-2}v^2(\delta_{\mu\nu}k^2-
 k_{\mu}k_{\nu})\exp\left[-i\frac{k^2}{4}(1-v^2)s\right].$$
 On inserting the values of the various integrals, and noticing that
 $$(\delta_{\mu\nu}k^2-k_{\mu}k_{\nu})A_{\mu}(-k)A_{\nu}(k)=\frac{1}{2}F_{\mu\nu}(-k)F_{\mu\nu}(k),$$
we obtain the immediately the gauge invariant form (with $s\rightarrow-is$)
 $$W^{(1)}=-\frac{e^2}{4\pi^2}\int d^4k\frac{1}{2}F_{\mu\nu}(-k)F_{\mu\nu}(k)\int_0^1 dv (1-v^2)\int_0^{\infty}ds s^{-1}\exp\left[-\left(m^2+\frac{k^2}{4}(1-v^2)\right)s\right]. $$
 This has been achieved without any special device, other than that of reserving the proper time integration to the last. A significant separation of terms is produced by a partial integration with respect to $v$, according to

 $$\int_0^1 dv (1-v^2)\int_0^{\infty}ds s^{-1}\exp[-(m^2+\frac{k^2}{4}(1-v^2))s]=\frac{2}{3}\int_0^{\infty}dss^{-1}\exp(-m^2s)-$$$$
 \frac{1}{2}k^2\int_0^1 dv (v^2-\frac{1}{3}v^4)\int_0^{\infty}ds \exp[-(m^2+\frac{k^2}{4}(1-v^2))s].$$
Adding the action integral of the Maxwell field, which is expressed in momentum space by

  $$W^{(0)}=-\int d^4k\frac{1}{4}F_{\mu\nu}(-k)F_{\mu\nu}(k),$$
 we obtain the modified integral,
 \b W=-\left[1+\frac{e^2}{12\pi^2}\int_0^{\infty}dss^{-1}\exp(-m^2s) \right]\int d^4k\frac{1}{4}F_{\mu\nu}(-k)F_{\mu\nu}(k)+$$$$
 \frac{e^2}{4\pi^2}\int d^4k\frac{k^2}{4}F_{\mu\nu}(-k)F_{\mu\nu}(k)\int_0^1 dv \frac{v^2(1-\frac{1}{3}v^2)}{m^2+\frac{k^2}{4}(1-v^2)}
 . \e
 The field strength and charge renormalization  produces the finite gauge invariant result.
  \b \label{eq:b} W=-\int d^4k\frac{1}{4}F_{\mu\nu}(-k)F_{\mu\nu}(k)\left[1- \frac{\alpha}{4\pi}\frac{k^2}{m^2}\int_0^1 dv\frac{v^2(1-\frac{1}{3}v^2)}{1+\frac{k^2}{4m^2}(1-v^2)}\right]
 . \e

\section{Calculations}
We briefly present some calculations and simplifications which has been used in this paper.

In this part we'll bring some calculations to simplify  the logarithmic functions which was used in section(4). It is easy to see that
$$K_+=\ln \left( 1+\frac{V}{k^2(k^2-m^2)-V}\right)=-\ln\left[1-\frac{V}{k^2(k^2-m^2)}\right].$$
Then, we can write
\begin{eqnarray}\ln\left(1-\frac{V}{k^2(k^2-m^2)}\right)&=&\ln\left[1+\frac{V}{m^2}(\frac{1}{k^2}-\frac{1}{k^2-m^2})\right]\nonumber\\
&=&\ln\left[\left(1+\frac{V}{m^2k^2}\right)\left(1-\frac{V}{m^2(k^2-m^2)}\right)+\frac{V^2}{m^4k^2(k^2-m^2)}\right],\end{eqnarray}
 and finally we get to
\b  K_+=-\ln\left(1+\frac{V}{m^2k^2}\right)-\ln\left(1-\frac{V}{m^2(k^2-m^2)}\right)-\ln\left(1+\frac{V^2}{(m^2k^2+V)(m^2k^2-m^4-V)}\right).\e
Now, we calculate
 \b K_-=\ln \left( 1-\frac{V}{k^2(k^2-m^2)-V}\right)=\ln \left( \frac{k^2(k^2-m^2)-2V}{k^2(k^2-m^2)-V}\right)$$
 $$=\ln \left( 1-\frac{2V}{k^2(k^2-m^2)}\right)- \ln \left( 1-\frac{V}{k^2(k^2-m^2)}\right).\e
So, we obtain
\begin{eqnarray} K_- &=&\ln\left(1+\frac{2V}{m^2k^2}\right)+\ln\left(1-\frac{2V}{m^2(k^2-m^2)}\right)\nonumber\\ & &+\ln\left(1+\frac{4V^2}{(m^2k^2+2V)(m^2k^2-m^4-2V)}\right)
 - \ln \left( 1-\frac{V}{k^2(k^2-m^2)}\right).\end{eqnarray}

 Now, we would like to present the calculation of $I_4$ in (4.4):
 \b I_4=-\int_0^1 dv \frac{v^2(1-\frac{1}{3}v^2)}{1-v^2}= -\frac{1}{2}\int_0^1 dv\left[ \frac{v^2(1-\frac{1}{3}v^2)}{1-v}+ \frac{v^2(1-\frac{1}{3}v^2)}{1+v}\right].\e
 By using the following relations:
 $$\frac{1}{1-v}=\sum_{n=0}^{\infty}v^n,\qquad \frac{1}{1+v}=\sum_{n=0}^{\infty}(-v)^n,$$
we obtain
\begin{eqnarray}I_4&=&-\frac{1}{2}\int_0^1 dv \sum_{n=0}^{\infty}\left[v^{n+2}-\frac{1}{3}v^{n+4}+ (-v)^{n+2}-\frac{1}{3}(-v)^{n+4}\right]
\nonumber\\ &=&-\frac{1}{2}\sum_{n=0}^{\infty} \left[\int_0^1 dvv^{n+2}-\frac{1}{3}\int_0^1 dvv^{n+4}+ \int_0^1 dv(-v)^{n+2}-
\frac{1}{3}\int_0^1 dv(-v)^{n+4}\right]\nonumber\\
&=&-\frac{1}{2}\sum_{n=0}^{\infty} \left[\frac{v^{n+3}}{n+3}-\frac{1}{3}\frac{v^{n+5}}{n+5}+ \frac{(-1)^nv^{n+3}}{n+3}-
\frac{1}{3} \frac{(-1)^nv^{n+5}}{n+5}\right]_{0}^{1}. \end{eqnarray}
And so, we can rewrite as
$$I_4=-\frac{1}{2}\left\{\left[-\ln(1-v)-v-\frac{v^2}{2}\right]+\frac{1}{3}\left[\ln(1-v)+v+\frac{v^2}{2}+\frac{v^3}{3}+\frac{v^4}{4}\right]
\right.$$$$\left.
+\left[\ln(1+v)-v+\frac{v^2}{2}\right]-\frac{1}{3}\left[\ln(1+v)-v+\frac{v^2}{2}-\frac{v^3}{3}+\frac{v^4}{4}\right]\right\}_{0}^{1} .$$
Finally, we have
\begin{eqnarray} I_4&=&-\left[-\frac{1}{3}\ln(1-v)+\frac{1}{3}\ln(1+v)-\frac{2}{3}v+\frac{1}{9}v^3\right]_{0}^{1}
\nonumber\\ &=&\frac{1}{3}\lim_{\mu\rightarrow0}\ln{\mu}-\frac{1}{3}\ln2+\frac{5}{9}.
 \end{eqnarray}

\end{appendix}

\end{document}